\documentclass[prd,nofootinbib,twocolumn,superscriptaddress,preprintnumbers,balancelastpage,nofootinbib]{revtex4-2}

\usepackage{graphicx}  
\usepackage{dcolumn}   
\usepackage{bm}        
\usepackage{amssymb,amsmath}
\usepackage{slashed}   
\usepackage[colorlinks=true,allcolors=black]{hyperref}
\usepackage{multirow}
\renewcommand{\eqref}[1]{Eq.~\ref{#1}}

\usepackage{dsfont}

\usepackage[usenames,dvipsnames,svgnames,table]{xcolor}

\def\be{\begin{equation}}
\def\ee{\end{equation}}
\def\bea{\begin{eqnarray}}
\def\eea{\end{eqnarray}}

\renewcommand{\vec}{\mathbf}

\definecolor{cborange}{HTML}{e69f00}
\definecolor{cbgreen}{HTML}{009e73}
\definecolor{cbyellow}{HTML}{f1dd42}
\definecolor{cblblue}{HTML}{56b4e9}
\definecolor{cbblue}{HTML}{0000FF}
\definecolor{defgrey}{HTML}{808080}
\definecolor{defgreen}{HTML}{008000}
\definecolor{defred}{HTML}{FA5F55}
\definecolor{darkred}{HTML}{9c240b}

\def\beq{\begin{equation}}
\def\eeq{\end{equation}}

\def\be{\begin{eqnarray}}
\def\ee{\end{eqnarray}}

\begin{document}
\widetext

\title{From 100 kpc to 10 Gpc: Dark Matter self-interactions before and after DESI}
\author{Salvatore Bottaro}\email{salvatoreb@tauex.tau.ac.il}
\affiliation{Raymond and Beverly Sackler School of Physics and Astronomy, Tel-Aviv 69978, Israel}
\author{Emanuele Castorina}\email{emanuele.castorina@unimi.it}
\affiliation{Dipartimento di Fisica “Aldo Pontremoli”, Universit\`a degli Studi di Milano and \\INFN, Sezione di Milano,
Via Celoria 16, 20133 Milan, Italy}
\author{Marco Costa}\email{mcosta1@perimeterinstitute.ca}
\affiliation{Perimeter Institute for Theoretical Physics, 31 Caroline St N, Waterloo, ON N2L 2Y5, Canada}
\author{Diego Redigolo}\email{diego.redigolo@fi.infn.it}
\affiliation{INFN, Sezione di Firenze,
Via Sansone 1, 50019 Sesto Fiorentino, Italy}
\author{Ennio Salvioni}\email{e.salvioni@sussex.ac.uk}
\affiliation{Department of Physics and Astronomy, University of Sussex, Sussex House, Brighton BN1 9RH, UK}

\begin{abstract}
We consider Dark Matter self-interactions mediated by ultralight scalars. We show that effectively massless mediators lead to an enhancement of the matter power spectrum, while heavier mediators lead to a suppression, together with a feature around their Jeans scale. We derive the strongest present constraints by combining Planck and BOSS data. The recent DESI measurements of Baryon Acoustic Oscillations exhibit a mild $2\hspace{0.2mm}\sigma$ preference for long-range self-interactions, as strong as 4 per mille of the gravitational coupling. Full-shape analyses of forthcoming DESI and Euclid data will confirm or disprove such a hint.
\end{abstract}

\maketitle
\noindent
\section{Introduction}

Cosmological observations on different scales have provided compelling evidence for the existence of Dark Matter (DM) and Dark Energy (DE), the first in the form of a collisionless non-relativistic fluid, and the second as a non-zero cosmological constant $\Lambda$ in its simplest incarnation. Together with the Standard Model (SM) of particle physics, these two fluids form the current standard description of the post-inflationary Universe, widely known as $\Lambda$CDM. 

The existence of the DM fluid is one of the most definitive indications of phenomena beyond the SM, making its characterization a central challenge in both particle physics and cosmology. In perfect analogy with the observed complex structure of the visible sector, it is likely that what we currently view as DM hides a rich Dark Sector (DS), whose dynamics simplifies to the one of a collisionless non-relativistic fluid on cosmological scales. Observing a deviation from $\Lambda$CDM could then provide crucial insights on the hidden microphysics of the DS.  

As observational precision improves, cosmology is uniquely positioned to test the properties of the DS directly, independent of any interactions connecting it to the SM. The exquisite knowledge of the Cosmic Microwave Background (CMB) provided by Planck~\cite{Planck:2018vyg} will soon be complemented by a comparable precision in measurements of galaxy clustering by the DESI survey first~\cite{DESI:2016fyo,DESI:2016igz}, the Euclid space telescope shortly thereafter~\cite{Euclid:2024yrr}, and new observational facilities proposed for the next decade~\cite{Schlegel:2019eqc,Schlegel:2022vrv,DESI:2024wkd,MSEScienceTeam:2019bva}.

Due to the need to describe perturbations beyond the linear regime, extracting full information from galaxy clustering data requires an advanced modelling provided by the Effective Field Theory of Large Scale Structure (EFTofLSS)~\cite{Baumann:2010tm,Carrasco:2012cv,Perko:2016puo,Foreman:2015lca,Carroll:2013oxa,Carrasco:2013mua,Pajer:2013jj,Senatore:2014vja,Angulo:2014tfa,Baldauf:2014qfa,Porto:2013qua,Vlah:2015sea,Chen:2020zjt}. The EFTofLSS has been successfully applied to present data~\cite{DAmico:2019fhj,Ivanov:2019pdj,DAmico:2020ods,Philcox:2020vvt,DAmico:2022osl,Ivanov:2023qzb,Chen:2021wdi} and is expected to provide unprecedented accuracy in the measurement of $\Lambda$CDM parameters when applied to future datasets~\cite{Sailer:2021yzm,Braganca:2023pcp}. In this article, we extend the EFTofLSS framework to the exploration of DM dynamics beyond the minimal $\Lambda$CDM model, focusing on the implications of DM self-interactions on cosmological scales.

For concreteness, we consider self-interactions mediated by an ultralight scalar field $\varphi$ with trilinear coupling to the DM $\chi$, $\mathcal{L}_{\text{int}} = -\, \kappa \varphi \chi^2$~\cite{Damour:1990tw,Wetterich:1994bg,Amendola:1999er,Farrar:2003uw,Archidiacono:2022iuu}. We expect this simple model to capture most of the phenomenology of attractive DM self-interactions in cosmological observables. Introducing the dimensionless field $s = G_s^{1/2} \varphi$, with $G_s\equiv\kappa^2/m_\chi^4\equiv4\pi G_N \beta$, we can write at the leading order in a $1/G_s$ expansion
\begin{equation}
-2\mathcal{L}\supset (\partial\chi)^2 + m_\chi^2(s)\chi^2 + \frac{1}{4\pi  G_N \beta}\left[(\partial s)^2 + m_\varphi^2 s^2\right] ,\label{eq:lag_2}
\end{equation}
which shows that the interaction can be traded for a DM space-time dependent mass, $m_\chi(s)=m_\chi (1+2s)^{1/2}$. The DM self-interaction is characterized by two parameters: its strength normalized to the one of gravity, $\beta$, and its range set by the Compton wavelength of the scalar, $\lambda_\varphi=1/m_\varphi$. As we show in Sec.~\ref{sec:masslessvsmassive}, the latter  is responsible for the modifications to the cosmological background. The imprints on the perturbations are instead controlled by the de Broglie wavelength of the scalar moving with the Hubble flow (\emph{i.e.}~the Jeans scale)~\cite{Hu:2000ke}, thus allowing cosmological observables to probe shorter ranges than naively expected. 

We focus on mediators with de Broglie wavelengths larger than the comoving horizon at matter-radiation equality, or equivalently Jeans wavenumbers $k_{\rm J}\lesssim 0.01\,h\,\mathrm{Mpc}^{-1}$, for which the EFTofLSS modelling developed in Ref.~\cite{Bottaro:2023wkd} is expected to hold. This implies that we can probe mediator masses as large as $m_\varphi \sim 10^5\, H_{0}\approx 10^{-28}\;\mathrm{eV}$. We show that only long-range forces with $\lambda_\varphi\geq H_0^{-1} \sim 4\,\mathrm{Gpc}$ lead to a scale-independent increase in the matter power spectrum compared to $\Lambda$CDM. For $\lambda_\varphi< H_0^{-1}$ the power spectrum is instead suppressed at any scale, with a step-like feature appearing around the Jeans scale of the massive mediator. An analytical explanation of these results is given in Sec.~\ref{sec:masslessvsmassive}. 

\begin{figure}
\includegraphics[width=0.5\textwidth]{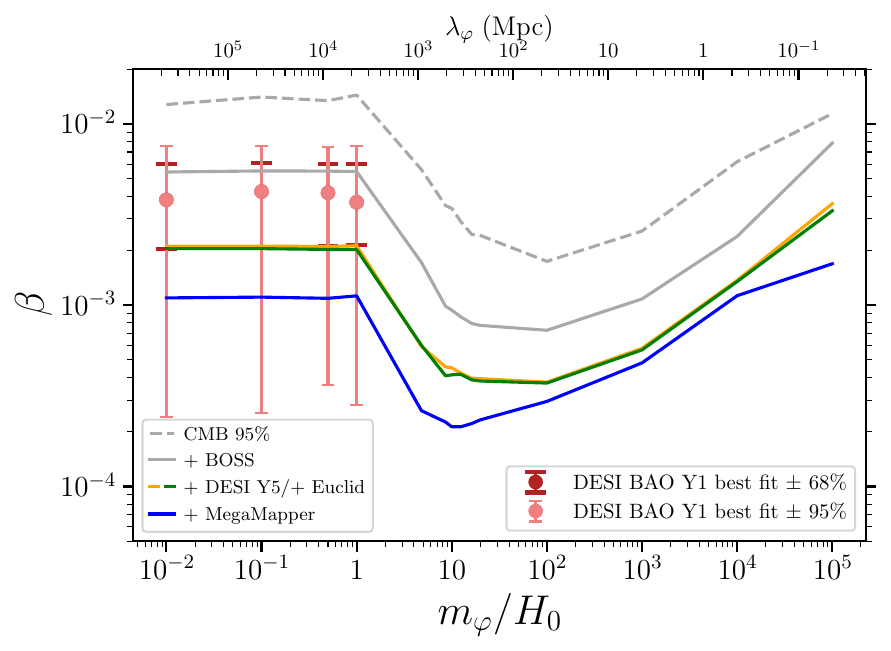}
\caption{Parameter space of DM self-interactions: $\beta
$ is the interaction strength relative to gravity and $(m_\varphi/ H_0)^{-1}$ sets the range of the interaction. The {\color{defgrey}\bf dashed gray line} is the $95\%$~CL constraint from Planck data~\cite{Planck:2018vyg}, the {\color{defgrey}\bf solid gray line} is the combination with BOSS full-shape data~\cite{BOSS:2016wmc} and previous BAO measurements~\cite{Kazin:2014qga,Beutler:2011hx,Ross:2014qpa}. The {\color{defred}\bf red dots} indicate the best fit points of the DESI BAO data~\cite{DESI:2024mwx} with a $95\%$ CL error band (the $68\%$ CL error band is shown in {\color{darkred}\bf darker red}). The {\color{cborange}\bf orange line} shows the  sensitivity of future DESI full-shape data~\cite{DESI:2016fyo,DESI:2016igz}, while the {\color{defgreen}\bf green line} corresponds to the forecasted reach of Euclid~\cite{EUCLID:2011zbd}. The {\color{cbblue}\bf blue line} gives the sensitivity of the proposed MegaMapper~\cite{Schlegel:2019eqc,Schlegel:2022vrv}. \label{fig:1d_bounds}}
\end{figure}

We obtain the current best limits on the coupling strength of the new force by combining the CMB with BOSS measurements~\cite{BOSS:2016wmc} of the full shape of the galaxy power spectrum, which we calculate using the EFTofLSS, and previous data on Baryon Acoustic Oscillations (BAO)~\cite{Kazin:2014qga,Beutler:2011hx,Ross:2014qpa}.\footnote{Replacing BOSS full-shape with BOSS BAO data gives nearly identical results, due to the limited BOSS statistics. This was already observed in Ref.~\cite{Archidiacono:2022iuu} for massless mediators.} 
These limits are shown as a gray solid line in Fig.~\ref{fig:1d_bounds}. For each value of $m_\varphi/H_0$, the interaction strength $\beta$ is scanned in a Markov Chain Monte Carlo (MCMC) analysis, simultaneously with the standard cosmological parameters and the additional parameters of the EFTofLSS. We then derive the projected sensitivities of the full-shape datasets of DESI, Euclid, and the proposed next-generation galaxy survey MegaMapper (colored lines in Fig.~\ref{fig:1d_bounds}). The details are presented in Sec.~\ref{sec:results}. The constraints and forecasts we report in Fig.~\ref{fig:1d_bounds} are the strongest to date in the whole window of DM self-interaction ranges considered.

Finally, we show that the recent DESI BAO measurement~\cite{DESI:2024mwx} seems to mildly favor long-range DM self-interactions over $\Lambda$CDM. The best fit points are shown in red in Fig.~\ref{fig:1d_bounds} and the physics driving this hint is discussed in Sec.~\ref{sec:DESI}. This intriguing preference will be conclusively confirmed or disproved by forthcoming full-shape data. 

\section{Massless vs massive mediators}\label{sec:masslessvsmassive}

In this section we discuss the main cosmological effects of DM self-interactions, presenting analytical results derived at leading order in an expansion for small $\beta$. The constraints in Fig.~\ref{fig:1d_bounds} justify this approximation. The background dynamics of the ultralight scalar is governed by its Klein-Gordon equation
\begin{equation}\label{eq:KG}
\bar{s}^{\prime\prime} + 2 \mathcal{H} \bar{s}^\prime + a^2 (m_\varphi^2 \bar{s} + 4\pi G_N \beta \widetilde{m}_s \bar{\rho}_\chi) = 0\,,
\end{equation}
where primes denote conformal time derivatives, $a$ is the scale factor, and $\widetilde{m}_s \equiv d\log m_\chi(s)/d s = (1 + 2\bar{s})^{-1}$. If the initial displacement of the scalar field is small, $\widetilde{m}_s \simeq 1$ at leading order in $\beta$. The scalar profile is sourced by the DM energy density, and its equation of state is dominated by its kinetic energy $(w_s \simeq +1$) as long as the mass term is negligible. 

During matter domination the energy transfer to the scalar accelerates the redshift evolution of the DM energy density, leading to a suppression of the Hubble parameter $H/H_{\Lambda\rm CDM} \simeq 1 - (\beta  \widetilde{m}_s^2 f_\chi^2/2) \log a / a_{\rm eq}\,$, where $f_\chi$ is the interacting fraction of the total matter (we define $f_x \equiv \bar{\rho}_x / \bar{\rho}_m$ for $x = \chi, b$ and, when relevant, $x = s$). The appearance of a large time logarithm is characteristic of the long-range nature of the interaction. If the mediator is effectively massless, $m_\varphi \leq H_0$, the above dynamics continues until today. Due to the suppression of the DM density, the comoving distance $\chi(z) = \int_0^z dz'/H(z')$ to any redshift ---including CMB last scattering--- is {\it increased}, at fixed matter content and Hubble rate, compared to the $\Lambda$CDM case. The present fraction of the energy density in the scalar, $f_s^{\rm massless} \simeq \beta \widetilde{m}_s^2 f_\chi^2/3$, is instead not log-enhanced and plays a subleading role in the phenomenology~\cite{Archidiacono:2022iuu}. 

For a massive mediator with $H_0 < m_\varphi \lesssim H_{\rm eq}\,$, the mass and DM source terms in~\eqref{eq:KG} become comparable at some time during matter domination, $a_{m_\varphi} \simeq  [ 3 H_{\rm eq} / (4\sqrt{2}\,m_\varphi)]^{2/3} a_{\rm eq}$. From this time on, the quadratic potential quickly dominates and the scalar begins to oscillate around the origin, eventually settling there. Its energy density redshifts like pressureless matter~\cite{Turner:1983he} and makes up a fraction of the matter density today that can be estimated as ($\log^2 x \equiv (\log x)^2$)
\begin{equation}\label{eq:fs}
f_{s} \simeq \frac{5}{4} f_s^{\rm massless}\, \log^2 \frac{H_{\rm eq}}{m_\varphi}\,,
\end{equation}
as long as $m_\varphi \ll H_{\rm eq}$. This enhanced contribution to the DM density at late times results in a {\it decrease} of the distances to all redshifts with respect to a $\Lambda$CDM model.

\begin{figure*}
\includegraphics[width=0.475\textwidth]{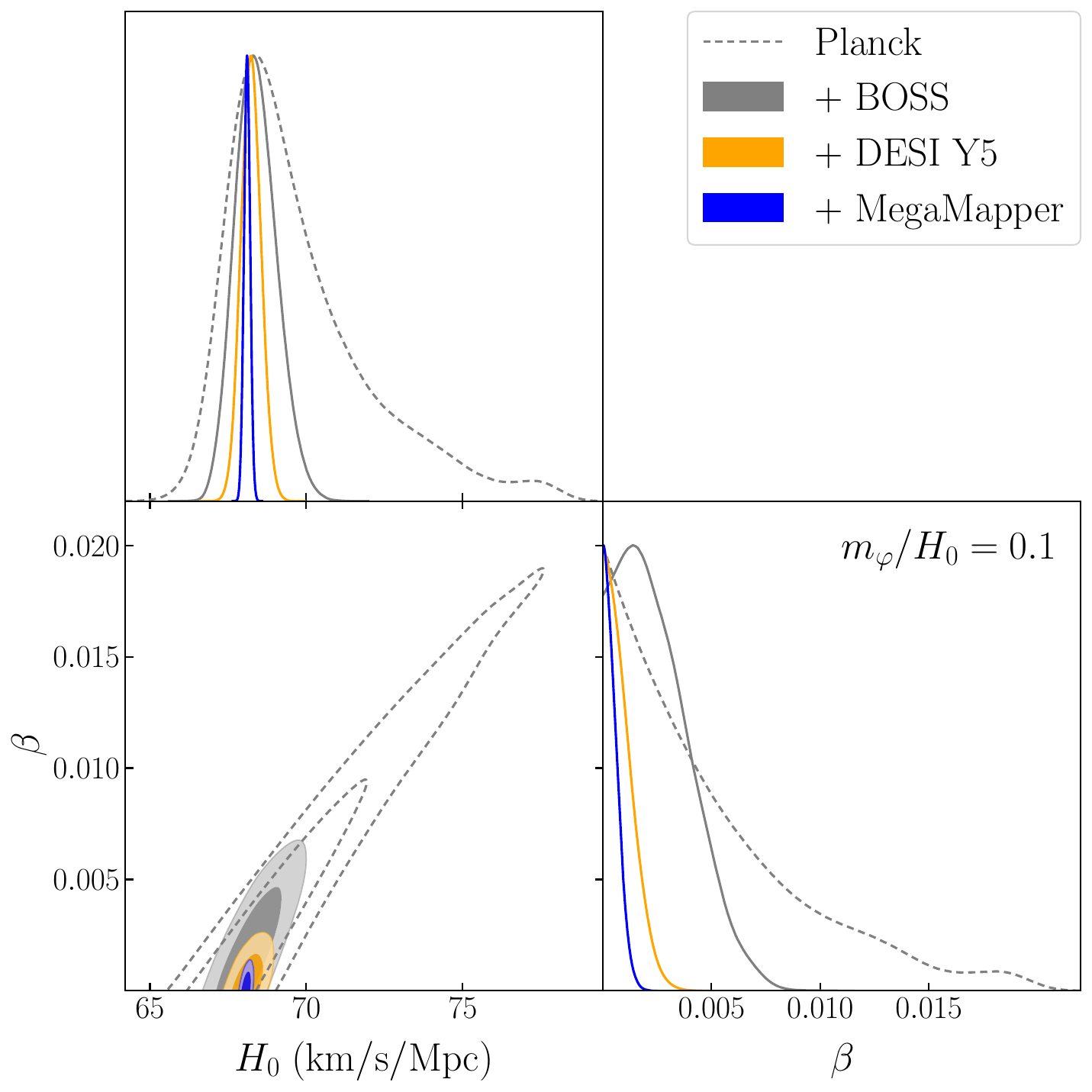}
\includegraphics[width=0.475\textwidth]{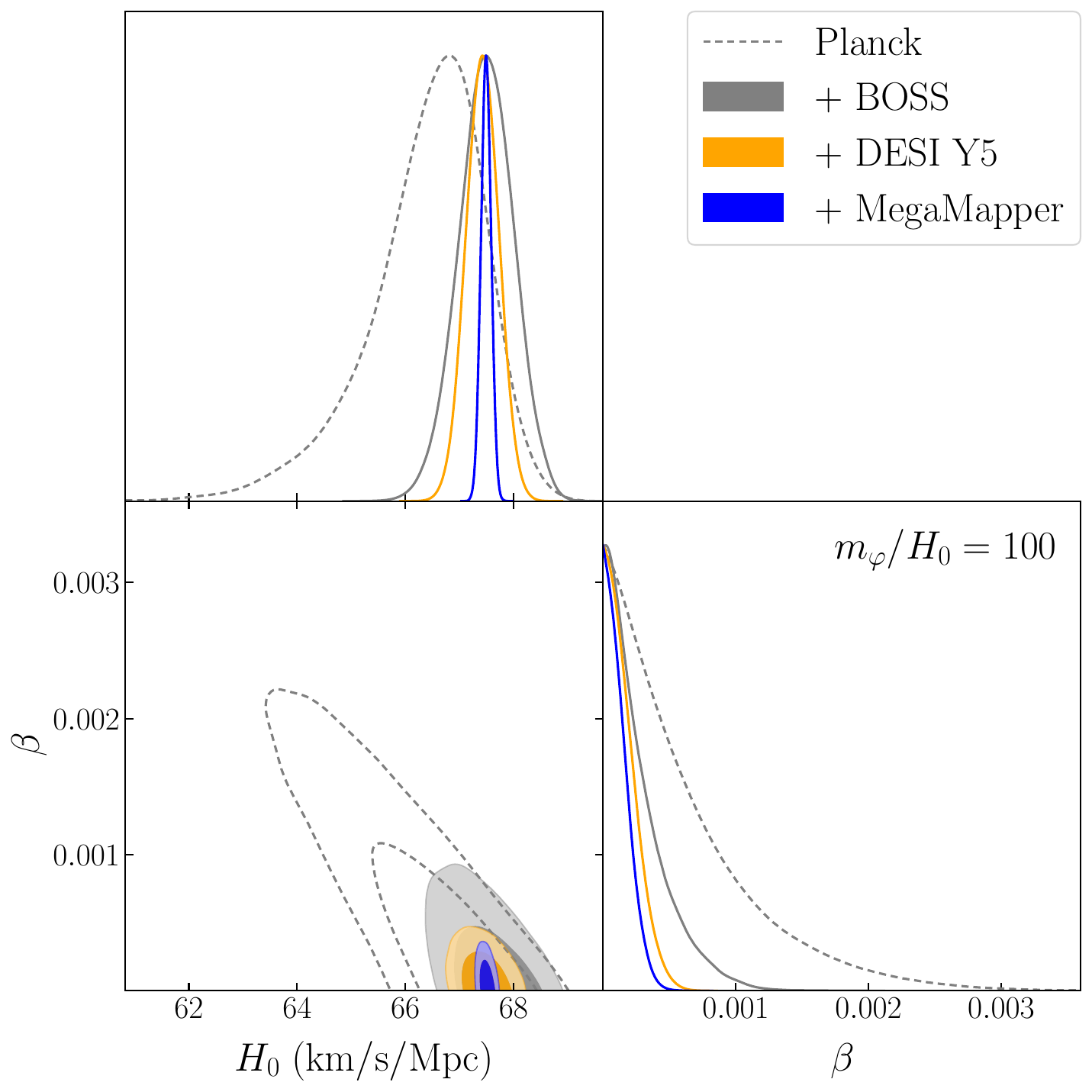}
\caption{Present and future constraints on the parameter space of DM self-interactions, at 68\% and 95\% CL. {\color{defgrey}\bf Gray dashed contours} correspond to Planck alone~\cite{Planck:2018vyg}, while the {\color{defgrey}\bf gray regions} show the combination with BOSS full-shape data~\cite{BOSS:2016wmc} and previous BAO measurements~\cite{Kazin:2014qga,Beutler:2011hx,Ross:2014qpa}. {\color{cborange}\bf Orange regions} correspond to our forecast for the full DESI dataset~\cite{DESI:2016fyo,DESI:2016igz}, while {\color{cbblue}\bf blue regions} show the reach of the proposed MegaMapper survey~\cite{Schlegel:2019eqc,Schlegel:2022vrv}. {\bf Left:} Effectively massless mediator with $m_\varphi<H_0$. {\bf Right:} Massive mediator with $m_\varphi/H_0=100$.}\label{fig:corner}
\end{figure*}

At the level of linear perturbations, $\delta_x \equiv \rho_x / \bar{\rho}_x - 1$, two main effects arise. First, there is a scale-independent enhancement of matter clustering caused by the attractive self-interaction, which operates as long as the mediator remains effectively massless~\cite{Archidiacono:2022iuu,Bottaro:2023wkd}. Second, once the mediator becomes a fraction of DM, its perturbations are strongly suppressed below its Jeans scale, $k \gg k_{\rm J} (a)$ with $k_{\rm J} \approx 3.9 \times 10^{-4}\, a^{1/4} \left( m_\varphi / H_0\right)^{1/2} \hspace{-0.5mm}(\Omega_m^0/ 0.3)^{1/4}\,h\,\mathrm{Mpc}^{-1}$~\cite{Hu:2000ke,Hlozek:2014lca}. This is a consequence of the macroscopic de Broglie wavelength of the ultralight scalar. Thus, for modes satisfying $k \gg k_{\rm J}$ at $a \gg a_{m_\varphi}$, the effect on the total matter density perturbation $\delta_m = f_\chi \delta_\chi + f_b \delta_b + f_s \delta_s$ is
\begin{equation}\label{eq:deltam}
\frac{\delta_m (a)}{\delta_m^{\rm CDM}(a)} - 1 \simeq  \frac{6}{5}\beta \widetilde{m}_s^2 f_\chi^2 \log \frac{a_{m_\varphi}}{a_{\rm eq}} - f_s - \frac{3}{5}f_s \log \frac{a}{a_{m_\varphi}} \,.
\end{equation}
The first modification to the CDM behavior corresponds to the scale-independent enhancement of growth active when $a < a_{m_\varphi}$, while the last two corrections embody the suppressed clustering of $s$ when $a > a_{m_\varphi}$. The $-f_s$ term simply manifests that a fraction $f_s$ of matter does not cluster, whereas the log-enhanced $f_s$ term encodes the impact on the $\chi$ plus baryon perturbations of the suppression of the gravitational potential by a factor $1 - f_s$. These effects are formally analogous to those of free-streaming massive neutrinos~\cite{Lesgourgues:2006nd}, although the underlying physical phenomena are distinct. For a massive mediator, the parametrically large $f_s$ (see~\eqref{eq:fs}) gives the dominant effect in~\eqref{eq:deltam}, causing an overall suppression of the linear power spectrum compared to $\Lambda$CDM and a step-like feature at $k\sim k_{\rm J}(a)$. Conversely, for a massless mediator one finds a nearly scale-independent increase of power, as previously shown in Refs.~\cite{Archidiacono:2022iuu,Bottaro:2023wkd}.

In this article we focus on masses $m_\varphi \lesssim H_{\rm eq}$, corresponding to $\lambda_\varphi \gtrsim 30\;\mathrm{kpc}$, which satisfy $k_{\rm J}(a) \lesssim k_{\rm eq} \equiv a_{\rm eq} H_{\rm eq}$. This ensures that $k\gtrsim k_{\rm J}$ for the modes of relevance to LSS, namely $k_{\rm eq} \lesssim k < k_{\rm NL}$ (where $k_{\rm NL}$ is the nonlinear scale). Therefore the EFTofLSS as derived for a massless mediator~\cite{Bottaro:2023wkd} still applies. Nonetheless, the theoretical uncertainties in our non-linear modelling increase toward the upper end of the considered mass window, $m_\varphi/H_0 \gtrsim 10^4$, because (i) eventually $k_{\rm J}$ becomes larger than $k_{\rm eq}$, hence a more careful treatment of the behavior of a relativistic fluid in the EFTofLSS is required; and (ii) the time logarithm $\log a_{m_\varphi}/a_{\rm eq}$ appearing in~\eqref{eq:deltam} is reduced, hence non-log-enhanced $\mathcal{O}(\beta)$ corrections become more important (see Appendix~\ref{app:massive} for further details).  

\section{Results}~\label{sec:results}

Current and projected bounds in the 2-dimensional parameter space of $(H_0,\beta)$ are presented in Fig.~\ref{fig:corner}. The left panel shows the case $m_\varphi/H_0 = 0.1$, while the right one corresponds to $m_\varphi/H_0 = 100$. Different colors indicate the combinations of present and future LSS datasets with Planck CMB measurements. The forecasts have been obtained using the \texttt{FishLSS} code~\cite{Sailer:2021yzm,FishLSS}, see Appendix~\ref{app:massive} for additional details. 

The degeneracies among the parameters can be easily understood. In the massless scenario, as discussed in Sec.~\ref{sec:masslessvsmassive}, the expansion rate is reduced at all times compared to the one in $\Lambda$CDM, which implies a larger distance to the last scattering surface (the physics prior to recombination is practically unchanged). The only way to recover the value of the angular size of the sound horizon as measured by Planck is therefore to increase $H_0$. As we impose flatness, this in turn drives the total matter fraction to lower values. 

The degeneracy is completely reversed in the massive scenario, as shown in the right panel of Fig.~\ref{fig:corner}. In this case, the Hubble rate exceeds the $\Lambda$CDM one after $a_{m_\varphi}$ because of the sizeable contribution of the mediator abundance to the late-time energy density, see~\eqref{eq:fs}. This results in a smaller distance to the last scattering surface, which is compensated by a lower value of $H_0$. 

We find that LSS datasets break the degeneracies in the CMB data, resulting in more than one order of magnitude improvement over Planck alone. The full DESI and Euclid datasets will achieve very similar bounds, while MegaMapper will push the sensitivity on $\beta$ below the per mille level for the whole mass range considered here. 

The shape of the bounds as functions of the mediator mass, shown in Fig.~\ref{fig:1d_bounds}, can also be qualitatively understood through the scaling in Eq.~\ref{eq:fs}. The bounds are strongest for $m_\varphi/H_0$ between $10$ and $10^3$, where $f_s$ is enhanced by more than one order of magnitude over its massless value. For larger masses, $f_s$ is less enhanced, thus resulting in weaker constraints. For smaller masses, we observe a rapid transition to the massless regime. Notice that the degeneracy directions are more elongated for massless than for massive mediators, as seen in Fig.~\ref{fig:corner}, which also impacts the final constraints.

Remarkably, the above results show that combining CMB and galaxy clustering data in scenarios beyond $\Lambda$CDM can lead to a much better sensitivity than naively expected, because of dynamical enhancements (here taking the form of large logs) and the presence of non-trivial correlations between the new physics and the $\Lambda$CDM parameters.

\section{Including DESI Year-1 BAO data}\label{sec:DESI}

\begin{figure}
\includegraphics[width=0.475\textwidth]{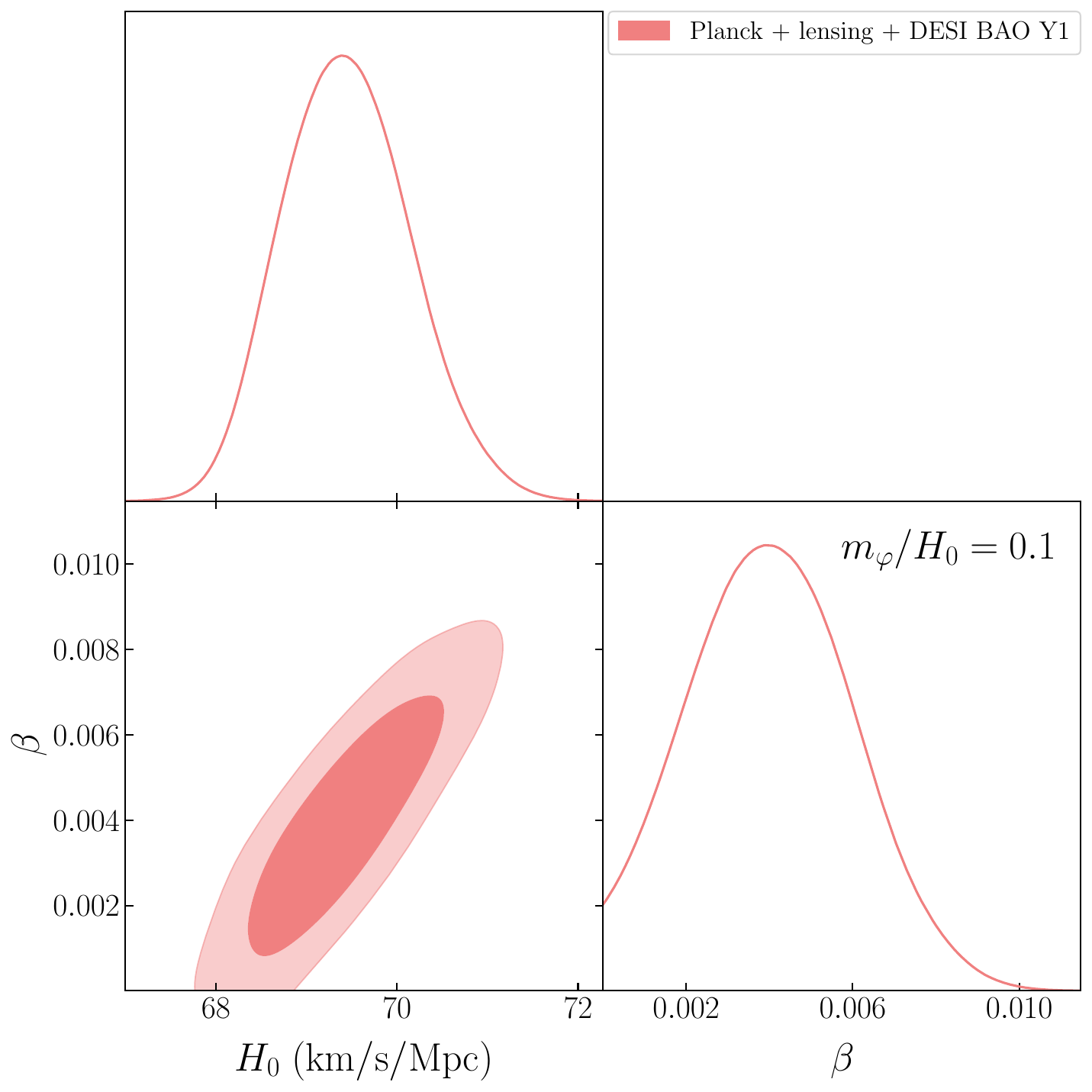}\hspace{10pt}
\caption{\label{fig:DESI}Fit to Planck CMB data, including measurements of the lensing potential~\cite{Planck:2018vyg,Planck:2018lbu}, combined with the recent DESI Year-1 BAO measurements~\cite{DESI:2024mwx}. An effectively massless mediator is assumed ($m_\varphi<H_0$).}
\end{figure}

Recently, the DESI collaboration released their first measurements of the BAO across a wide range of redshifts and with several different tracers~\cite{DESI:2024mwx,DESI:2024lzq,DESI:2024uvr}. The BAO technique measures the quantities $H(z)r_d $ and $D_M(z)/r_d$, where $r_d$ is the sound horizon at the baryon drag epoch and $D_M(z)$ is the transverse comoving distance (equivalent to $\chi(z)$ for spatially flat models), thus providing a unique sensitivity to the background dynamics of the Universe at low redshifts. In particular, the new DESI measurements of $H r_d$  at $z\sim 0.5$ and of $D_M/r_d$ at $z\sim 0.7$, both with Luminous Red Galaxies (LRG), are a few $\sigma$'s different than the prediction of a Planck best fit $\Lambda$CDM cosmology. In combination with Type Ia Supernovae, these findings have been interpreted as a possible evidence for an evolving Dark Energy component ~\cite{DESI:2024mwx,DESI:2024kob,Gialamas:2024lyw,Notari:2024rti}. Figure~\ref{fig:DESI} shows that, intriguingly, the low-$z$ DESI measurements can be partially accommodated in a scenario where DM experiences a long-range self-interaction with $\beta \approx 0.004\, \pm\, 0.002$. This is due to the higher values of $H_0$ allowed in our model, as shown in Figs.~\ref{fig:corner} and \ref{fig:DESI} (see also Appendix~\ref{sec:DESifit} for more details). However, it should be kept in mind that some DESI data points for LRG appear to be in tension with previous, and equally accurate, BOSS BAO measurements at the same redshifts. Future DESI and Euclid data releases are thus needed to shed light on these results.  

In our minimal extension of $\Lambda$CDM, the non-zero value of $\beta$ preferred by Planck plus DESI BAO is in mild tension with supernova measurements, as shown in Appendix~\ref{sec:DESifit}. The situation could in principle be improved if the same scalar mediating the long-range self-interaction played the role of quintessence at late times. 

Finally, long-range DM self-interactions have been recently advocated as a possible explanation of the long-standing  hint for an enhancement of late-time clustering in CMB lensing measurements~\cite{Craig:2024tky}. As already mentioned in Ref.~\cite{Archidiacono:2022iuu}, a consistent treatment of the background cosmology and the dynamics of matter fluctuations leads to a large cancellation in the lensing potential, effectively diminishing the imprint of long-range self-interactions. In Appendix~\ref{sec:DESifit} we detail a simple analytic argument for this cancellation. For reference, our best fit point in Fig.~\ref{fig:DESI} increases the lensing potential by less than $2\%$.

\section{Outlook}\label{sec:outlook}

In this article, we showed how galaxy clustering data reveals powerful insights on DM self-interactions, independent of any couplings with the visible sector. We presented world-leading cosmological constraints on attractive self-interactions with range $\lambda_\varphi \gtrsim H_{\text{eq}}^{-1} \sim 30$ kpc, derived from our analysis of the BOSS galaxy power spectrum within the EFTofLSS, combined with Planck CMB data and BAO measurements. Forthcoming galaxy clustering data will improve the sensitivity by around one order of magnitude, exploring a vast and largely uncharted parameter space of DM self-interactions.

The EFTofLSS allows us to consistently incorporate new physics parameters alongside those that encode our partial understanding of the clustering (bias coefficients and ultraviolet counterterms, see Appendix~\ref{sec:DESifit}). By fitting all these parameters simultaneously to data, we enhance the robustness of the derived constraints against theoretical systematics. For the ranges considered here, precision cosmology outperforms complementary astrophysical probes, establishing itself as a powerful tool for understanding DM self-interactions~\cite{Kesden:2006vz,Desmond:2020gzn,Bogorad:2023wzn}.

Remarkably, the distinctive features of massive mediators enable the power spectrum to differentiate between self-interactions of different ranges. As an example of this discriminating potential, we showed that the recent Year-1 DESI BAO measurements signal a mild preference for long-range self-interactions, which would accommodate the higher value of $H_0$ observed by DESI. Future full-shape data from DESI, along with data from Euclid, will be able to fully confirm or disprove such a hint. 

The possibility for galaxy clustering data to probe interaction ranges shorter than $H_{\text{eq}}^{-1}$ is currently limited by the lack of a consistent theoretical modelling. Developing such an extension of the EFTofLSS will allow us to test DM self-interactions up to $k_{\rm J} \sim k_{\text{NL}}$, corresponding to ranges as short as $1\text{ kpc}$~\cite{relativisticpaper}. Probing even smaller ranges will most likely require to go beyond the EFTofLSS~\cite{KVTtoappear}. Moreover, scenarios where only a fraction of the DM energy density is self-interacting (\emph{i.e.}~$f_\chi\ll 1$) will change the imprints in the power spectrum in an interesting way, enhancing the effects of relative density and velocity fluctuations over the ones of background modifications. We plan to come back to this possibility in the near future~\cite{FractionPaper}, completing the picture of what precision cosmology can teach us about DM self-interactions.

\vspace{1cm}
\acknowledgments
The numerical results presented in this work made extensive use of public software including \texttt{CLASS}~\cite{Blas:2011rf,Lesgourgues:2011re}, \texttt{MontePython}~\cite{Audren:2012wb,Brinckmann:2018cvx}, \texttt{GetDist}~\cite{Lewis:2019xzd} and \texttt{PyBird}~\cite{DAmico:2020kxu,PyBird}. The modified version of \texttt{CLASS} used here was developed in Ref.~\cite{Archidiacono:2022iuu}, for which we thank Maria Archidiacono. We are grateful to Alessio Notari, Michele Redi and Andrea Tesi for useful discussions and for sharing with us their likelihood for the DESI BAO data. We thank Cyril Creque-Sarbinowski, Olivier Simon and Ken Van Tilburg for discussions about their related upcoming work~\cite{KVTtoappear}. SB is supported by the Israel Academy of Sciences and Humanities \& Council for Higher Education Excellence Fellowship Program for International Postdoctoral Researchers. MC is supported in part by Perimeter Institute for Theoretical Physics. Research at Perimeter Institute is supported by the Government of Canada through the Department of Innovation, Science and Economic Development Canada and by the Province of Ontario through the Ministry of Research, Innovation and Science. The work of DR is supported in part by the European Union - Next Generation EU through the PRIN2022 Grant n.~202289JEW4. ES is supported by the Science and Technology Facilities Council under the Ernest Rutherford Fellowship ST/X003612/1, and thanks the CERN Theory Department for generous hospitality during the completion of this project. The work of MC, DR and ES was performed in part at the Aspen Center for Physics, which is supported by National Science Foundation grant PHY-2210452.

\bibliographystyle{JHEP}
\bibliography{5thLSS.bib}

\clearpage
\newpage
\appendix
\onecolumngrid
\clearpage
\newpage
\appendix
\onecolumngrid
\section{Massive vs massless forces in detail}\label{app:massive}

We begin with a discussion of the background cosmology. For $a < a_{m_\varphi}$ we solve the Klein-Gordon (KG) equation in~\eqref{eq:KG} neglecting the mass term, which, switching to $y \equiv a/a_{\rm eq}$ as time variable, reads
\begin{equation}
y^2 \frac{H^2}{H_{\rm eq}^2} \frac{\partial^2 \bar{s}}{\partial y^2} + \left( 4y \frac{H^2}{H_{\rm eq}^2} + y^2 \frac{H}{H_{\rm eq}} \frac{\partial}{\partial y} \frac{H}{H_{\rm eq}} \right) \frac{\partial \bar{s}}{\partial y} + \frac{3}{4} \beta \widetilde{m}_s f_\chi  y^{-3} = 0\,,
\end{equation}
where $H / H_{\rm eq} = y^{-3/2}  \sqrt{1 + y^{-1}}/\sqrt{2}\,$. This admits the solution
\begin{equation} \label{eq:s_early}
\bar{s} (y)_{\rm early} = \bar{s}_{\rm ini} + 2 \beta \widetilde{m}_s f_\chi  \left( - \log \frac{\sqrt{1+y} + 1}{2} - \frac{1}{2} + \frac{\sqrt{1+y} - 1}{y}   \right) \stackrel{y \gg 1}{\simeq} - \beta \widetilde{m}_s f_\chi  \log y\,.
\end{equation}
Making use of this result, the time $y_{m_\varphi}$ when the mass term and DM source term in~\eqref{eq:KG} become equal satisfies $y_{m_\varphi}^3 \log y_{m_\varphi} = 3 / (4 \eta_\varphi^2)$, where $\eta_{\varphi} \equiv m_\varphi / H_{\rm eq}$. The solution is
\begin{equation}\label{eq:oscMatch}
  y_{m_\varphi} = \frac{1}{\mathcal{G}(\eta_\varphi)} \Big( \frac{3}{4\sqrt{2}\eta_\varphi} \Big)^{2/3}\,, \qquad \mathrm{where}\qquad \mathcal{G}(\eta_\varphi) \equiv \frac{1}{2}\, W_0^{1/3} \Big(\frac{9}{4\eta_\varphi^2}\Big)\,,
\end{equation}
with $W_0(x)$ denoting the product logarithm, which can be approximated as $W_0(x)\simeq \log x-\log(\log x)$ in the relevant range of $\eta_\varphi$. Numerically, $\mathcal{G}(\eta_\varphi)$ is an $\mathcal{O}(1)$ function which slowly decreases from $1.38$ at $\eta_\varphi = 10^{-5}$ to $0.48$ at $\eta_\varphi= 1$. 

For $a > a_{m_\varphi}$ it is a good approximation to neglect in the KG equation the DM source term, which rapidly becomes negligible. In this limit, the KG equation can be written as 
\begin{equation}\label{eq:s_late}
\frac{1}{2y} \frac{\partial^2 \bar{s}}{\partial y^2} + \frac{5}{4y^2} \frac{\partial \bar{s}}{\partial y} + \eta_\varphi^2 \bar{s} = 0 \quad \to \quad \bar{s}(y)_{\rm late} =  \frac{\beta \widetilde{m}_s f_\chi }{\eta_\varphi  y^{3/2}}\,   \bigg[ A(\eta_\varphi) \cos \bigg( \frac{2\sqrt{2}}{3} \eta_\varphi\, y^{3/2} \bigg) + B(\eta_\varphi) \sin \bigg( \frac{2\sqrt{2}}{3} \eta_\varphi\, y^{3/2} \bigg) \bigg]\,,
\end{equation}
where matter domination was assumed, and $A,B$ are $\mathcal{O}(1)$ coefficients determined by matching at $y = y_{m_\varphi}$ with the early-time solution in~\eqref{eq:s_early}. Equation~\ref{eq:s_late} allows us to obtain an expression for the late-time energy density fraction in the scalar 
\begin{equation}\label{eq:Omega_s}
\Omega_s (y \gg y_{m_\varphi}) \simeq \frac{\beta  \widetilde{m}_s^2 f_\chi^2 H_{\rm eq}^2}{6 H_0^2 y^{3}}  \left[ \frac{9}{4} \Big( 1 + \frac{1}{4\, \mathcal{G}^3(\eta_\varphi)} \Big) \log^2 \frac{\big(\frac{3}{4\sqrt{2}\eta_\varphi}\big)^{2/3}}{\mathcal{G}(\eta_\varphi)} + 3 \log \frac{\big(\frac{3}{4\sqrt{2}\eta_\varphi}\big)^{2/3}}{\mathcal{G}(\eta_\varphi)}  + 1 \right]\,,
\end{equation}
where we defined $\Omega_i \equiv \bar{\rho}_i / \bar{\rho}_{\rm cr}$ for the $i$ species, with $\bar{\rho}_{\rm cr} = 3H_0^2 / (8\pi G_N)$.

For $m_\varphi \ll H_{\rm eq}$ it is a good approximation to retain only the leading $\sim \log^2 \eta_\varphi$ term in the above formula, yielding
\begin{equation}\label{eq:fs_limit}
f_s = \frac{\Omega_s}{\Omega_\chi + \Omega_b + \Omega_s} \stackrel{y \gg y_{m_\varphi}}{\simeq} f_s^{\rm massless} \Big( 1 + \frac{1}{4\, \mathcal{G}^3(\eta_\varphi)} \Big) \log^2 \eta_\varphi\,,
\end{equation}
where $f_s^{\rm massless} = \beta \widetilde{m}_s^2 f_\chi^2/3$ is the result one finds~\cite{Archidiacono:2022iuu} for $m_\varphi\leq H_0\,$, by employing the solution in~\eqref{eq:s_early}. When providing analytical results in the main text, such as~\eqref{eq:fs}, we have taken $\mathcal{G} = 1$ as representative value. Notice that in this work and in Refs.~\cite{Archidiacono:2022iuu,Bottaro:2023wkd} we defined $\Omega_i = \bar{\rho}_i/\bar{\rho}_{\rm cr}$ for the individual species but $\Omega_m = \bar{\rho}_m/\bar{\rho}_{\rm tot}(a)$, hence $\Omega_\chi + \Omega_b + \Omega_s \neq \Omega_m$ in~\eqref{eq:fs_limit}.  
\begin{figure}[t]
\includegraphics[width=0.46\textwidth]{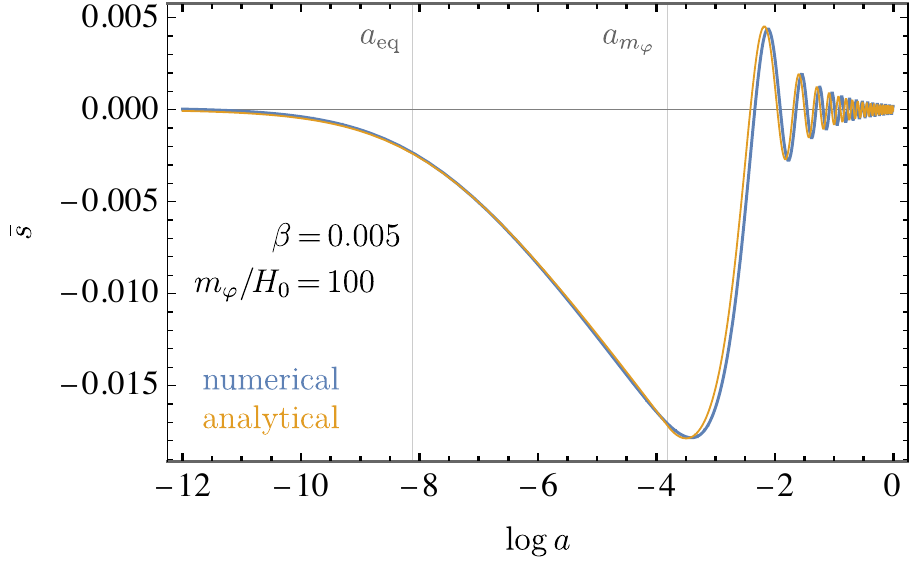}\hspace{3mm}
\includegraphics[width=0.457\textwidth]{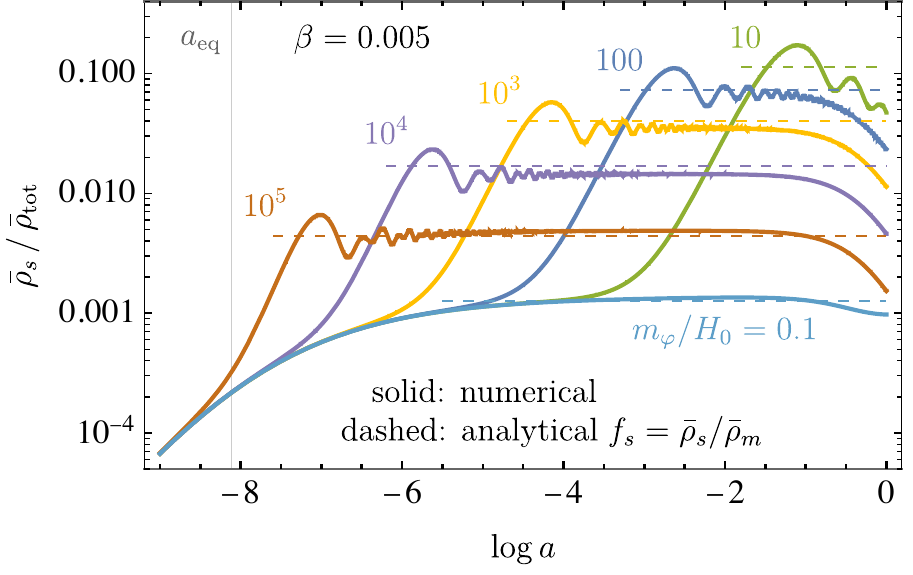}
\caption{\label{fig:bkg}Time evolution of background quantities and comparison with our analytical solutions derived up to $\mathcal{O}(\beta)$. {\bf Left:} Background value of the scalar field, $\bar{s}$. The analytical solution is based on Eqs.~\ref{eq:s_early} and~\ref{eq:s_late}. {\bf Right:} Fraction of the total energy density in the scalar field. Constant dashed lines show the analytical results for $f_s$, which for the massive scenario are based on~\eqref{eq:Omega_s}. In both panels we have set $\widetilde{\Omega}_d = 0.27$ (where $\widetilde{\Omega}_d$, defined as the fractional energy density $\chi$ would have today had it evolved like $a^{-3}$~\cite{Archidiacono:2022iuu}, is a proxy for $\Omega_\chi^0$) and $h = 0.67$.}
\end{figure}
\begin{figure}[t]
\includegraphics[width=0.45\textwidth]{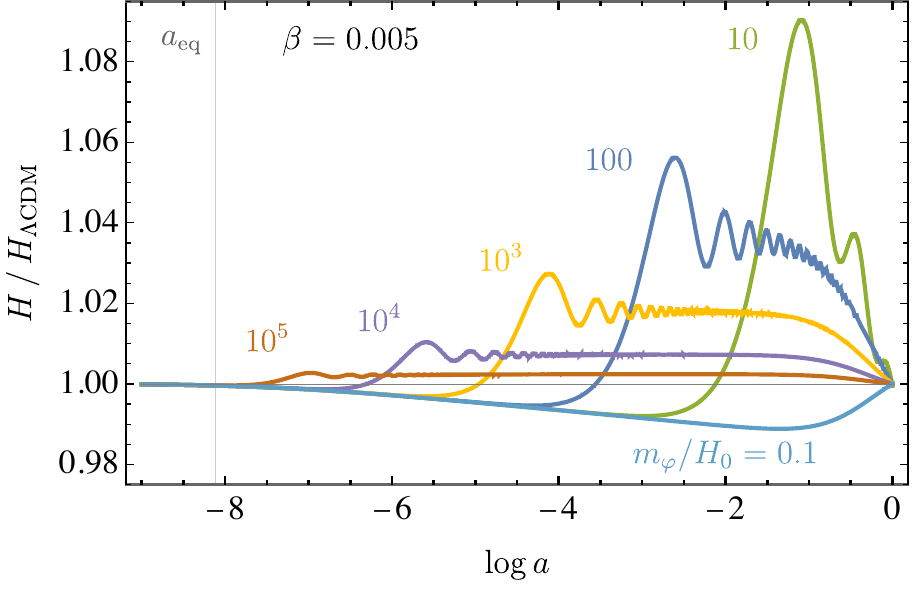}\hspace{4mm}\includegraphics[width=0.45\textwidth]{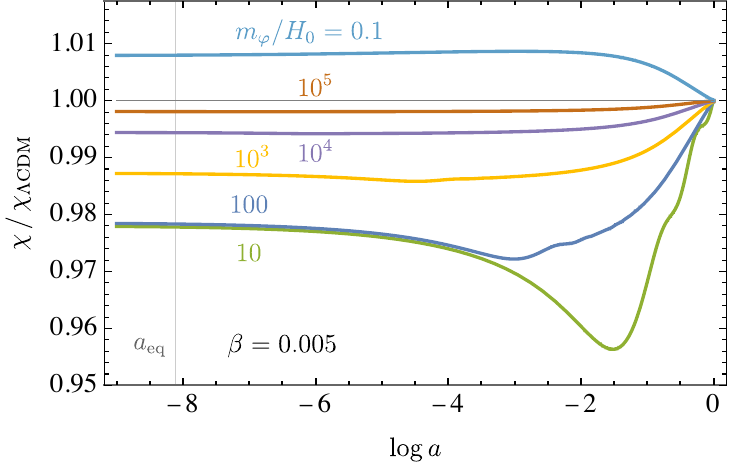}
\caption{\label{fig:bkg2}Ratios of background quantities to their values in $\Lambda$CDM. {\bf Left:} Hubble parameter. {\bf Right:} Comoving distance. In both panels we have set $\widetilde{\Omega}_d = 0.27$ and $h = 0.67$.}
\end{figure}

\begin{figure}[t]
\includegraphics[width=0.45\textwidth]{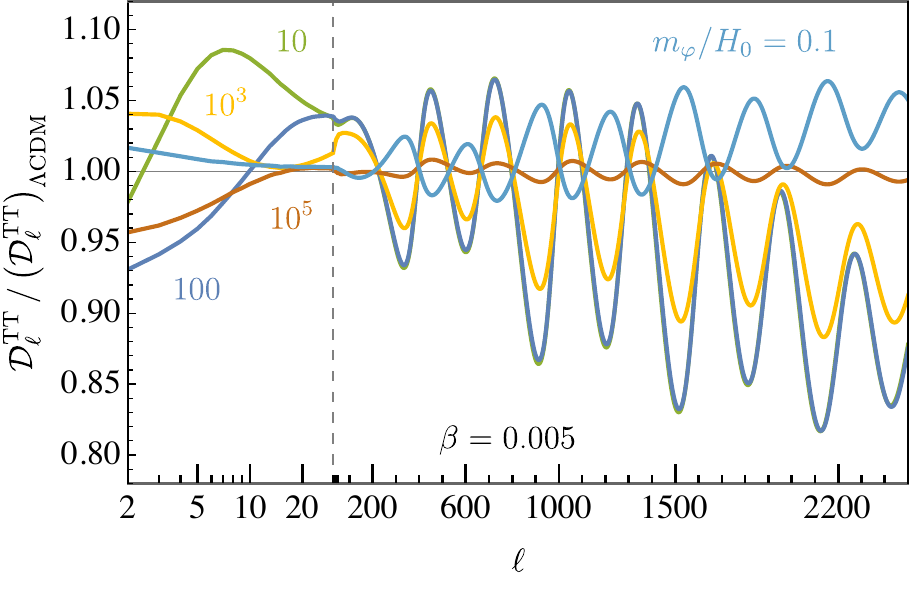}\hspace{3mm}
\includegraphics[width=0.449\textwidth]{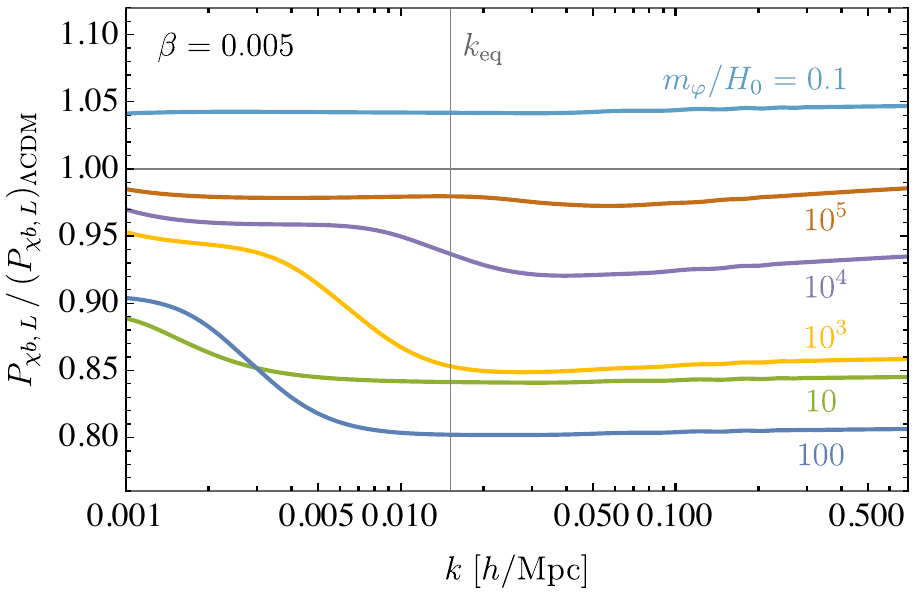}
\caption{\label{fig:pert}{\bf Left:} Ratio of CMB temperature power spectra to $\Lambda$CDM. The scale on the horizontal axis switches from logarithmic to linear at $\ell = 30$. We do not show $m_\varphi/H_0 = 10^4$ to avoid clutter. {\bf Right:} Ratio of the linear power spectrum for the $\chi$-$b$ fluid to $\Lambda$CDM. The vertical line indicates $k_{\rm eq} \equiv \mathcal{H}_{\rm eq}$. In both panels we have set $\widetilde{\Omega}_d = 0.27$ and $h = 0.67$.}
\end{figure}

The evolution of key background quantities is shown in Figs.~\ref{fig:bkg} and~\ref{fig:bkg2}, fixing $\beta = 0.005$ for illustration. In the left panel of Fig.~\ref{fig:bkg}, the analytical solution for $\bar{s}$ up to $\mathcal{O}(\beta)$ presented in Eqs.~\ref{eq:s_early} and~\ref{eq:s_late} is found to reproduce accurately the numerics. In the right panel we show the strong enhancement of the scalar energy density in the massive case. Again, our analytical result in~\eqref{eq:Omega_s} provides a good approximation. The left panel of Fig.~\ref{fig:bkg2} shows the effect on the Hubble parameter, making it evident that for any mass $m_\varphi > H_0$ the late-time enhancement of $H$ due to $\bar{\rho}_s$ dominates over the earlier suppression caused by $\bar{\rho}_\chi < \bar{\rho}_{\rm CDM}$. This results in a suppressed comoving distance, $\chi (z) < \chi_{\Lambda \rm CDM} (z)$, for any $z$, as shown in the right panel of the same figure.

Finally we observe that, for $m_\varphi > H_0$, after a few oscillations the dynamics of $\bar{s}$ becomes identical to that of a decoupled ultralight scalar field which was initially frozen to an appropriately chosen initial condition $\bar{\varphi}_{\rm ini} < 0$, and begins oscillating at $a_{\rm osc} \simeq  [ 3 H_{\rm eq} / (\sqrt{2}\,m_\varphi)]^{2/3} a_{\rm eq}$ (corresponding to $3H \simeq m_\varphi$; notice that $ a_{\rm osc} > a_{m_{\varphi}}$ by an $\mathcal{O}(1)$ factor)~\cite{Hlozek:2014lca}.

\vspace{0.2cm}
We then turn to the cosmological perturbations. In the left panel of Fig.~\ref{fig:pert} we show the ratio to $\Lambda$CDM of the CMB temperature power spectrum, setting again $\beta = 0.005$ for all mediator masses. The most important effect of DM self-interactions on the CMB is a shift in the locations of the acoustic peaks and troughs, as originally noticed in Ref.~\cite{Archidiacono:2022iuu} for massless mediators. This is the result of two facts:~(1) physical scales at recombination are mostly unaffected, because the interaction effectively turns on only around matter-radiation equality; (2) the distance to last scattering is modified compared to $\Lambda$CDM, as shown in the right panel of Fig.~\ref{fig:bkg2}. The shifts have opposite phases for massless and massive mediators, as a consequence of the distance being increased or decreased, respectively, relative to $\Lambda$CDM. At low $\ell$ we also observe corrections to the integrated Sachs-Wolfe effect, as the self-interaction causes the gravitational potentials to vary in time. The largest impact happens at very low redshift, when the cosmological constant takes over, corresponding to very large scales. Notice that the temperature power spectra for $m_\varphi/ H_0 = 10$ and $100$ are almost identical for $\ell \gtrsim 30$: this is due to the distances to any $z \gtrsim 50$ being accidentally very similar for those two masses, as the right panel of Fig.~\ref{fig:bkg2} demonstrates.

In the right panel of Fig.~\ref{fig:pert} we show the ratio to $\Lambda$CDM of the linear power spectrum of the $\chi$-baryon density perturbation, $\delta_{\chi b} \equiv (\bar{\rho}_\chi \delta_\chi + \bar{\rho}_b \delta_b) / (\bar{\rho}_\chi + \bar{\rho}_b)$. For a massive mediator, this perturbation grows at $a \gg a_{m_\varphi}$ and $k \gg k_{\rm J}$ as
\begin{equation} \label{eq:log_terms}
\delta_{\chi b} (a) \simeq \Big( 1 + \frac{6}{5}\beta \widetilde{m}_s^2 f_\chi^2 \log \frac{a_{m_\varphi}}{a_{\rm eq}} - \frac{3}{5}f_s \log \frac{a}{a_{m_\varphi}} \Big) \delta_{\chi b}^{\rm CDM} (a)\,,
\end{equation}
which is related to the total matter perturbation in~\eqref{eq:deltam} by $\delta_m = (1 - f_s) \delta_{\chi b} + f_s \delta_s$. As demonstrated in Refs.~\cite{Villaescusa-Navarro:2013pva,Castorina:2013wga} for the analogous case of massive neutrinos, $\delta_{\chi b}$ is the appropriate perturbation to utilize in the galaxy bias expansion, where the non-clustering field needs to be omitted. As discussed in Sec.~\ref{sec:masslessvsmassive}, in this work we focus on mediator masses satisfying $m_\varphi \lesssim H_{\rm eq}$, or equivalently $m_\varphi/H_0 \lesssim 1.5 \times 10^5$, because this ensures that the Jeans wavenumber~\cite{Hu:2000ke,Hlozek:2014lca}
\begin{equation}\label{eq:Jeans}
k_{\rm J}(a) \approx 3.9 \times 10^{-4}\, a^{1/4} \left( \frac{ m_\varphi }{  H_0}\right)^{1/2} \Big(\frac{\Omega_m^0}{ 0.3 }\Big)^{1/4}\,h\;\mathrm{Mpc}^{-1}
\end{equation}
is below the window $k_{\rm eq} \lesssim k \lesssim k_{\rm NL}$ relevant for galaxy surveys (we define $k_{\rm eq} = \mathcal{H}_{\rm eq}$, with $\mathcal{H} \equiv a'/a$). It is thus justified to perform the perturbative calculation of galaxy correlators by means of the EFTofLSS for massless mediators, which was presented in Ref.~\cite{Bottaro:2023wkd}. For example, the $1$-loop correction to the real space galaxy power spectrum in Eulerian perturbation theory reads 
\begin{equation} \label{eq:SPT_example}
\Delta P_{g,{\rm 1\mbox{-}loop}}(k, a) = 2 \int \frac{\mathrm{d}^3 p}{(2\pi)^3} \Big[ F_{2,g}(\vec{p}, \vec{k-p}) \Big]^2 P_{\chi b,L}(p)  P_{\chi b,L} (|\vec{k}-\vec{p}|) + 6\, b_1 P_{\chi b, L}(k) \int \frac{\mathrm{d}^3 p}{(2\pi)^3} F_{3,g}( \vec{p}, - \vec{p}, \vec{k} ) P_{\chi b, L}(p)\,,
\end{equation}
where all linear power spectra are evaluated at $a$. The Einstein-de Sitter galaxy kernels, $F_{n,g}$, are defined as
\begin{align}
F_{2,g}(\vec{k}_1, \vec{k}_2)&= b_1 F_2 (\vec{k}_1, \vec{k}_2) + \frac{b_2}{2} + b_{K^2} \Big(\frac{(\vec{k}_1 \cdot \vec{k}_2)^2}{k_1^2 k_2^2} - \frac{1}{3} \Big)\,,\\
F_{3,g}(\vec{k}_1, \vec{k}_2,\vec{k}_3)&= b_1 F_3 (\vec{k}_1, \vec{k}_2, \vec{k}_3 ) + \frac{1}{3!}\sum_{\rm perm} \bigg[ b_2 F_2 (\vec{k}_2, \vec{k}_3) +2 b_{K^2} \left( \frac{(\vec{k}_1 \cdot \vec{k}_{23})^2}{k_1^2 k_{23}^2} - \frac{1}{3} \right) F_2 (\vec{k}_2, \vec{k}_3)\notag \\
&\hspace{4.5cm} +\, \frac{4}{7} b_{\Gamma_3} \left( \frac{(\vec{k}_1 \cdot \vec{k}_{23})^2}{k_1^2 k_{23}^2} - 1 \right)\left( 1 - \frac{(\vec{k}_2 \cdot \vec{k}_3)^2}{k_2^2 k_3^2} \right)\bigg]\,,
\end{align}
in terms of the bias coefficients $b_i$ (see Ref.~\cite{Bottaro:2023wkd} for our conventions). The standard matter kernels $F_n$ can be found in Ref.~\cite{Bernardeau:2001qr}. A single counterterm proportional to $k^2 P_{\chi b, L}(k)$ is sufficient to absorb the ultraviolet sensitivity of the kernels. The expressions above account for the leading log-enhanced corrections shown in~\eqref{eq:log_terms}, while they neglect the effects of relative perturbations between $\chi$ and $b$, whose growth is not enhanced by large logarithms (results for the relative perturbations were given in Ref.~\cite{Bottaro:2023wkd}). In practice the code we use to perform our Fisher forecasts, \texttt{FishLSS}~\cite{Sailer:2021yzm,FishLSS}, employs redshift space and relies on Lagrangian perturbation theory. Nevertheless, the simplified example above suffices to illustrate the most important features of the treatment we apply.

We stress, however, that towards the upper end of our mass window -- namely, for $m_\varphi/H_0 \gtrsim 10^4$ -- improvements to the EFTofLSS modelling are expected to be necessary. These modifications have two distinct motivations:

\begin{enumerate}
\item[(i)] The Jeans wavenumber in~\eqref{eq:Jeans} actually becomes larger than $k_{\rm eq}$ for $m_\varphi/H_0 \gtrsim 10^4$, as it can also be observed in the right panel of Fig.~\ref{fig:pert}. As the new physics scale $k_{\rm J}$ falls in the window $k_{\rm eq} \lesssim k \lesssim k_{\rm NL}$, the factorization of space- and time-dependences that holds to high accuracy for the $\Lambda$CDM perturbative solutions breaks down. As a consequence, we expect the appearance of non-trivial time integrals over $\tilde{a} < a$ on the right-hand side of~\eqref{eq:SPT_example}, together with new corrections to the standard kernels and contributions of $s$ to the clustering of matter at scales $k\lesssim k_{\rm J}$. 
\item[(ii)] As the mediator mass increases, the time logarithm that accounts for the effectively massless phase in~\eqref{eq:log_terms}, $\log a_{m_\varphi}/a_{\rm eq}$, becomes smaller and ultimately of $\mathcal{O}(1)$. Therefore, the non-log-enhanced terms associated with relative fluctuations become comparatively more important and need to be included for an accurate treatment. For reference, in the case of massless mediators the error associated to neglecting relative perturbations was estimated to be of order $1/(f_\chi \log a/a_{\rm eq}) \approx 10\%/f_\chi$, where $a$ corresponds to the redshift of the observed galaxy sample~\cite{Bottaro:2023wkd}.
\end{enumerate}
The above effects cannot be consistently included with currently available tools. We plan to do so in future publications~\cite{relativisticpaper,FractionPaper}, where the EFTofLSS will be extended to even heavier mediators and the effects of relative fluctuations will be consistently included. Nonetheless, concerning (i) we note that for $m_\varphi/H_0 \gtrsim 10^4$ the fraction $f_s$, which controls the amplitude of the step-like feature in the linear power spectrum, is less strongly enhanced. As a consequence, we expect that corrections of type (i) to the analysis presented in this work will be moderate.

Lastly, we provide some technical details about our forecasts, which have been derived using \texttt{FishLSS}~\cite{Sailer:2021yzm,FishLSS}. We take the default samples with the exception of DESI, where we add the Bright Galaxy Survey to the default Emission Line Galaxies (ELG). As already explained, the galaxy bias expansion is written in terms of the $\chi$-baryon field, as the scalar field does not cluster on the relevant scales \cite{Castorina:2013wga,Villaescusa-Navarro:2013pva}. To obtain the forecasts we first invert the covariance matrix of the MCMC chains from Planck+BOSS data to estimate the Fisher matrix. Next, we add it to the Fisher matrix of the forecasts obtained with \texttt{FishLSS}. We then generate a Gaussian sample centered at the MCMC best fit value (setting $\beta$ to 0) with covariance given by the inverse of the combined Fisher matrix. Finally, we drop all points with $\beta <0$ and obtain the 68\% and 95\% CL intervals.

\section{Further details on the fit to DESI Year-1 BAO data}\label{sec:DESifit}

\begin{figure}[h!]
\includegraphics[width=0.45\textwidth]{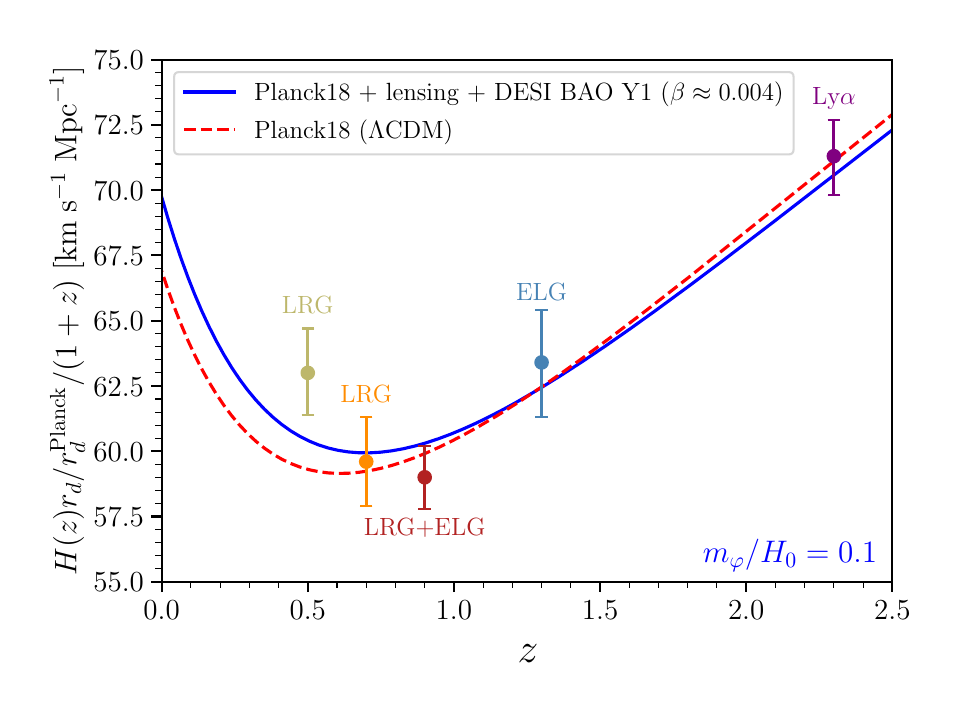}\hspace{2mm}
\includegraphics[width=0.449\textwidth]{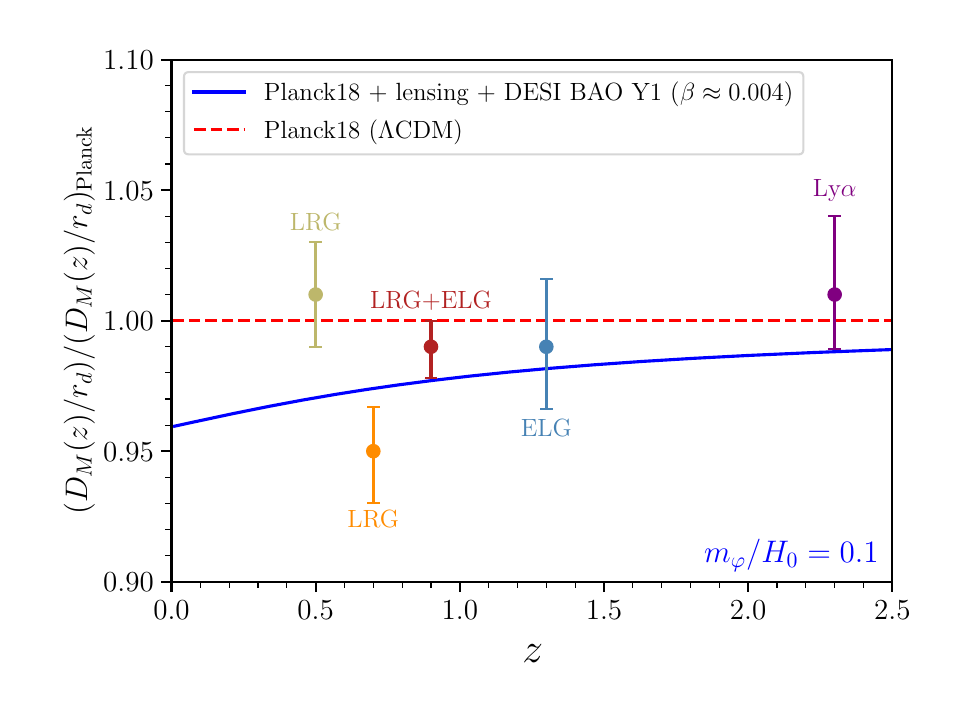}
\caption{\label{fig:DESI_details} Comparison of the DESI BAO measurements for different tracers (colored points)~\cite{DESI:2024mwx} to the predictions of our Planck+lensing+DESI BAO Y1 best fit cosmology with long-range DM self-interactions {\color{cbblue}\bf (solid blue lines)} and the predictions of a Planck best fit $\Lambda$CDM cosmology {\color{red}\bf (dashed red lines)}~\cite{Planck:2018vyg}. {\bf Left:} $H(z)r_d$, suitably normalized. {\bf Right:} $D_M(z)/r_d$, normalized to the Planck best fit value.}
\end{figure}

Here we provide further information on the mild preference of the DESI BAO Y1 data for a long-range DM self-interaction with $m_\varphi \leq H_0$ over $\Lambda$CDM. In Fig.~\ref{fig:DESI_details} we show the predictions of our best fit point with $\beta \approx 0.004$ (see Fig.~\ref{fig:DESI}) for the BAO observables $H(z) r_d$ and $D_M(z)/r_d$, comparing them to the predictions of a best fit $\Lambda$CDM cosmology for Planck alone~\cite{Planck:2018vyg} and to the DESI data points for different tracers~\cite{DESI:2024mwx}. Turning on a long-range DM self-interaction visibly improves the agreement with the two measurements that are about $2\sigma$ away from the $\Lambda$CDM predictions, namely $H r_d$ at $z \sim 0.5$ and $D_M/r_d$ at $z \sim 0.7$, both obtained with LRG. However, in our minimal extension of $\Lambda$CDM there is a residual tension with the DESI measurement of $D_M/r_d$ at $z \sim 0.5$, which agrees very well with the $\Lambda$CDM prediction. This issue may be ameliorated in scenarios where the scalar mediator also plays the role of a quintessence field, thus modifying the dynamics at very low redshift, along the lines discussed in Refs.~\cite{DESI:2024mwx,DESI:2024kob,Gialamas:2024lyw,Notari:2024rti}.

\subsection*{Impact of lensing and supernova data for massless mediators}

As mentioned in Sec.~\ref{sec:DESI}, the impact of DM self-interactions mediated by a massless scalar on the CMB lensing power spectrum is rather mild, owing to a significant cancellation between the suppression of the total matter density and the enhancement of clustering, which was already observed in Ref.~\cite{Archidiacono:2022iuu} (see Fig.~8 there). Here we provide an analytical understanding of this cancellation. The angular power spectrum of the lensing potential $\phi$ can be expressed as an integral of the power spectrum of the gravitational potential, $\langle\Psi(\vec{k},\tau)\Psi(\vec{k}',\tau)\rangle=(2\pi)^3\delta(\vec{k}+\vec{k}')P_{\Psi}(k,\tau)$. In the Limber approximation, valid for $L\gtrsim 20$, one finds~\cite{Lewis:2006fu}

\begin{equation}\label{eq:lenspot}
    C_L^{\phi\phi}=4\int_0^{\chi_*}\frac{\mathrm{d}\chi}{\chi^2}P_{\Psi}\left(\frac{L}{\chi},\tau_0-\chi\right)\left(\frac{\chi_*-\chi}{\chi_*\chi}\right)^2 \simeq \frac{9}{L^4(\tau_0 - \tau_\ast)^2}\int_{\tau_\ast}^{\tau_0}\mathrm{d}\tau  \mathcal{H}^4 (\tau)\,P_m\left(\frac{L}{\tau_0 - \tau},\tau\right)(\tau_\ast - \tau)^2\,,
\end{equation}
where $\chi = \tau_0 - \tau$ is the comoving distance and $\ast$ denotes quantities evaluated at recombination. In the second equality we have used the Poisson equation to replace the power spectrum of $\Psi$ with that of matter, and set $\Omega_m(\tau)\simeq 1$ because the integral is dominated by the matter epoch. The integral in~\eqref{eq:lenspot} can be evaluated perturbatively in $\beta$ by making use of results presented in Refs.~\cite{Archidiacono:2022iuu,Bottaro:2023wkd},
\begin{equation}
\mathcal{H}(\tau) \simeq \frac{2}{\tau}(1 - \varepsilon f_\chi)\,,\qquad P_{m,L} (k, \tau) \simeq \Big( 1 + \frac{4}{5}\varepsilon f_\chi \log \frac{\tau}{\tau_{\rm eq}}- \frac{4}{25}\varepsilon f_\chi \Big) P_{m,L}^{\rm CDM}(k, \tau)\,, \qquad \varepsilon \equiv \beta \widetilde{m}_s^2 f_\chi\,,
\end{equation}
as well as approximating $P_{m,L}^{\rm CDM}(k, \tau) = \mathcal{P} \tau^4 k^{-n}$, where $n \approx 1.5$ in the relevant region of wavenumbers and $\mathcal{P}$ collects all the $k$- and $\tau$-independent factors. We perform the integration and employ the relation between scale factor and conformal time, $a(\tau) \simeq \mathcal{C}\tau^2 \big(1 - 2\varepsilon f_\chi \log \tau/ \tau_{\rm eq} + 7 \varepsilon f_\chi / 3 \big)$ with $\mathcal{C} = \Omega_m^0 H_0^2 / 4$, arriving at
\begin{equation}
\frac{C_L^{\phi\phi}}{(C_L^{\phi\phi})_{\rm CDM}} \simeq 1 + \frac{33}{20} \varepsilon f_\chi \log \frac{a_0}{a_{\rm eq}} - 7.6\, \varepsilon f_\chi \approx 1 + (13.4 -  7.6) \varepsilon f_\chi \approx  1 + 5.8\, \varepsilon f_\chi \,,
\end{equation}
where in the second equality we have taken $z_{\rm eq} \approx 3400$. Thus the leading correction to the lensing power spectrum is an $L$-independent enhancement, but a significant cancellation is at play. It is useful to compare the above result with those for other relevant quantities, e.g.~the linear matter power spectrum,
\begin{figure}[b]
\includegraphics[width=0.475\textwidth]{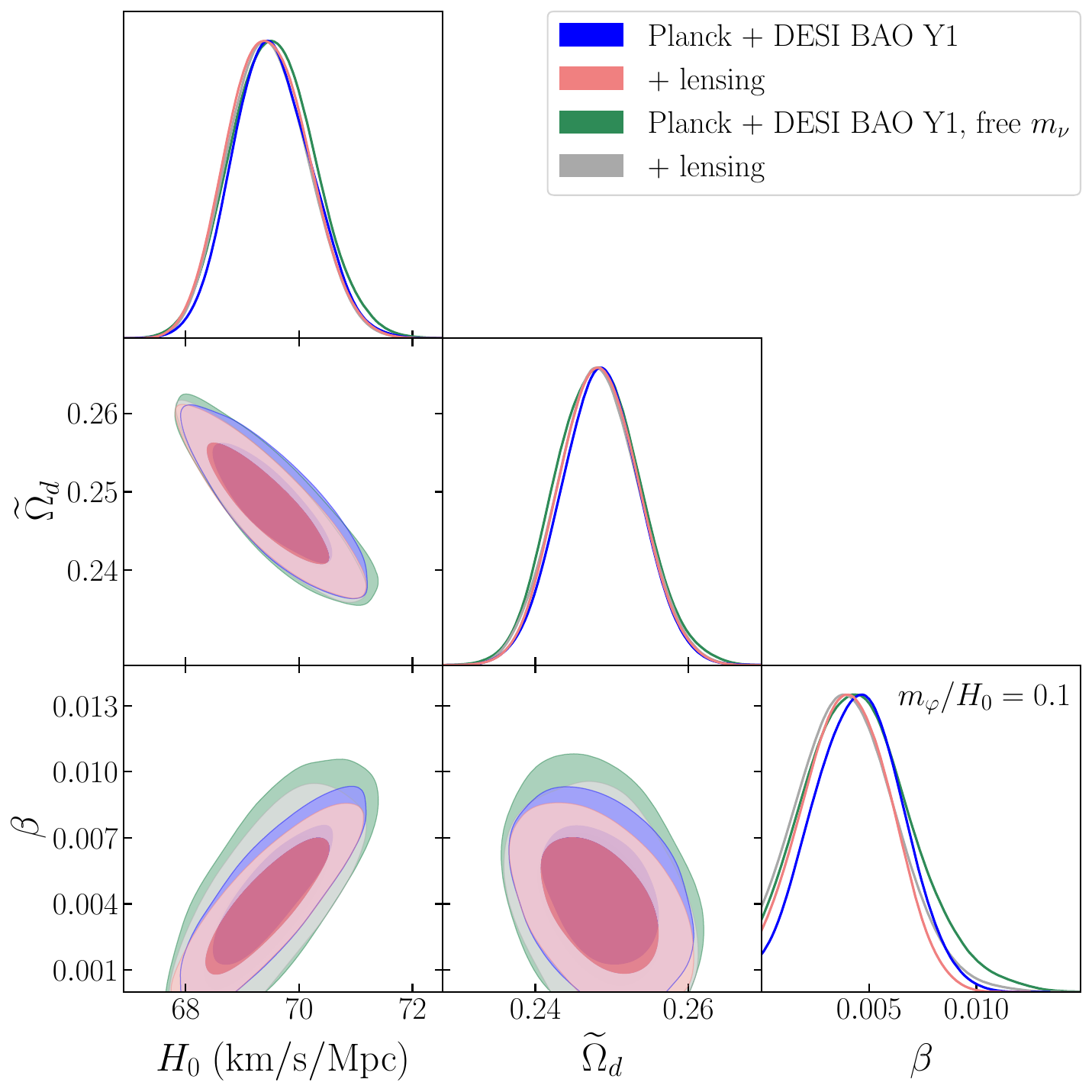}\hfill
\includegraphics[width=0.475\textwidth]{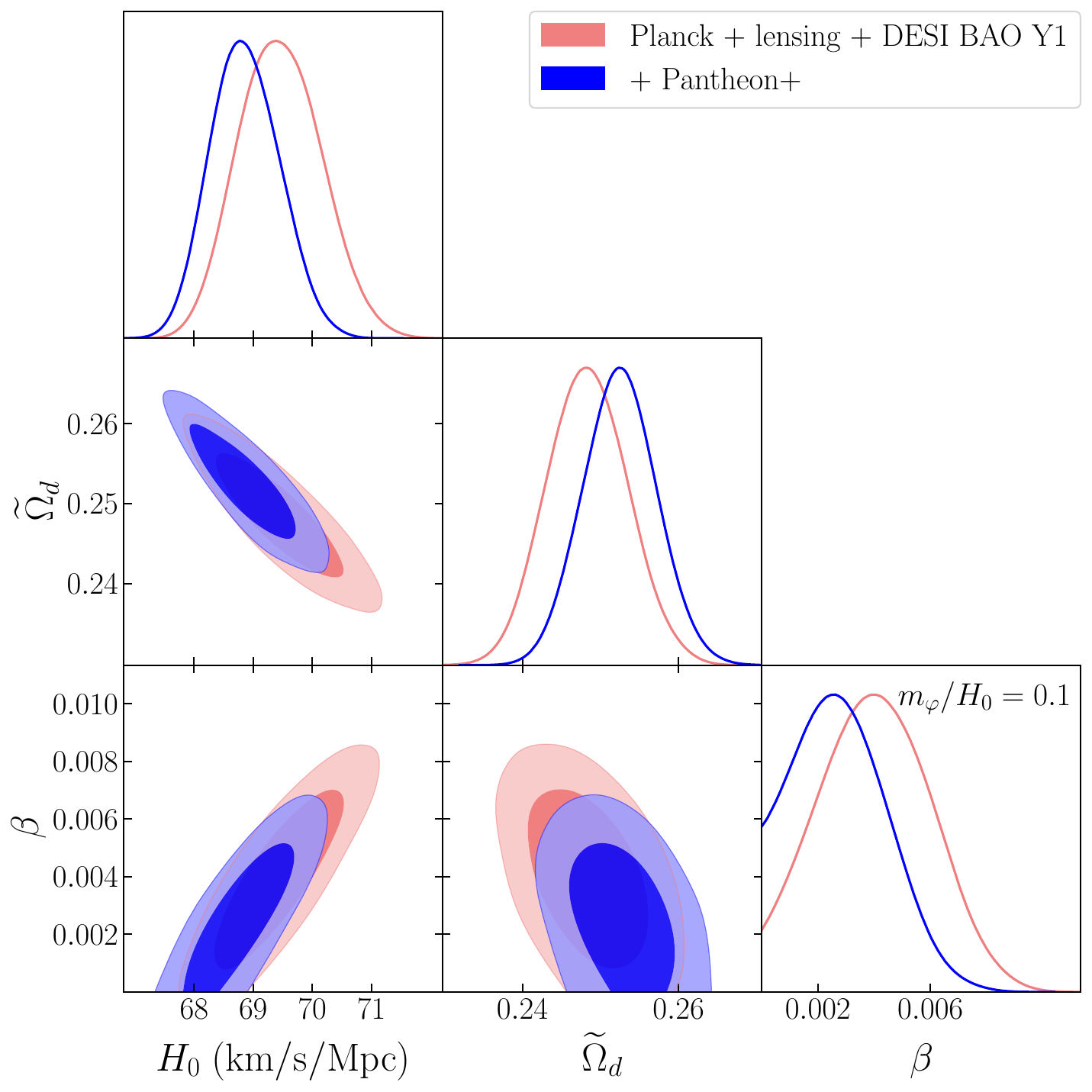}
\caption{{\bf Left:} The impact on the DESI fit in Fig.~\ref{fig:DESI} of including Planck CMB lensing data~\cite{Planck:2018lbu} and of varying the neutrino masses. See Sec.~\ref{sec:DESI} for further comments. {\bf Right:} The impact of Pantheon+ supernova data~\cite{Scolnic:2021amr}.\label{fig:appB}}
\end{figure}
\begin{equation}
\frac{P_{m,L} }{ P_{m,L}^{\rm CDM}} \simeq 1 + \frac{12}{5}  \varepsilon f_\chi \Big( \log \frac{a_0}{a_{\rm eq}} - \frac{181}{90} \Big) \approx 1 + ( 19.5 - 4.8) \varepsilon f_\chi \approx 1 + 14.7\, \varepsilon f_\chi    \,,
\end{equation}
which indeed displays a stronger sensitivity to the coupling $\beta$ of the self-interaction. In particular, the DESI best fit point shown in the left panel of Fig.~\ref{fig:DESI} leads to a scale-independent enhancement of less than $2\%$ for the lensing potential and approximately $4.5\%$ for the matter power spectrum. The left panel of Fig.~\ref{fig:appB} confirms the small impact of adding Planck CMB lensing data~\cite{Planck:2018lbu} in the fit and also indicates a weak dependence on the neutrino masses.

Finally, in the right panel of Fig.~\ref{fig:appB} we show the impact of the Pantheon+ supernova dataset~\cite{Scolnic:2021amr} on our model. At face value, since the dynamics during Dark Energy domination is left unchanged with respect to $\Lambda$CDM, the Pantheon+ dataset points toward a smaller value of $H_0$ (a higher value of $\Omega_m^0$) compared to Planck, disfavoring long-range DM self-interactions.   

\end{document}